%% file: 0_main.tex
\documentclass[sigconf]{acmart}

\copyrightyear{2025}
\acmYear{2025}
\setcopyright{cc}
\setcctype{by-nc}
\acmConference[SIGIR '25]{Proceedings of the 48th International ACM SIGIR Conference on Research and Development in Information Retrieval}{July 13--18, 2025}{Padua, Italy}
\acmBooktitle{Proceedings of the 48th International ACM SIGIR Conference on Research and Development in Information Retrieval (SIGIR '25), July 13--18, 2025, Padua, Italy}\acmDOI{10.1145/3726302.3729973}
\acmISBN{979-8-4007-1592-1/2025/07}

\AtBeginDocument{%
  }

\usepackage{algorithm}
\usepackage{algorithmic}
\usepackage{subfigure}
\usepackage{multirow}
\usepackage{balance}
\begin{document}

\title{Exploring Training and Inference Scaling Laws in Generative Retrieval}

\author{Hongru Cai}
\email{henry.hongrucai@gmail.com}
\affiliation{
\institution{National University of Singapore}
\country{Singapore}}

\author{Yongqi Li}
\authornote{Corresponding authors.}
\email{liyongqi0@gmail.com}
\affiliation{
\institution{The Hong Kong Polytechnic University}
\city{Hong Kong SAR}
\country{China}}

\author{Ruifeng Yuan}
\email{ruifeng.yuan@connect.polyu.hk}
\affiliation{
\institution{The Hong Kong Polytechnic University}
\city{Hong Kong SAR}
\country{China}}

\author{Wenjie Wang}
\email{wenjiewang96@gmail.com}
\authornotemark[1]
\affiliation{
\institution{University of Science and Technology of China}
\city{Hefei}
\country{China}}

\author{Zhen Zhang}
\email{cristinzhang7@gmail.com}
\affiliation{
\institution{Nanyang Technological University}
\country{Singapore}}

\author{Wenjie Li}
\email{cswjli@comp.polyu.edu.hk}
\affiliation{
\institution{The Hong Kong Polytechnic University}
\city{Hong Kong SAR}
\country{China}}

\author{Tat-Seng Chua}
\email{dcscts@nus.edu.sg}
\affiliation{
\institution{National University of Singapore}
\country{Singapore}}

\renewcommand{\shortauthors}{Hongru Cai et al.}

\begin{abstract}
Generative retrieval reformulates retrieval as an autoregressive generation task, where large language models (LLMs) generate target documents directly from a query. As a novel paradigm, the mechanisms that underpin its performance and scalability remain largely unexplored. We systematically investigate \textbf{training and inference scaling laws} in generative retrieval, exploring how model size, training data scale, and inference-time compute jointly influence performance. We propose a novel evaluation metric inspired by contrastive entropy and generation loss, providing a continuous performance signal that enables robust comparisons across diverse generative retrieval methods. Our experiments show that n-gram-based methods align strongly with training and inference scaling laws. We find that increasing model size, training data scale, and inference-time compute all contribute to improved performance, highlighting the complementary roles of these factors in enhancing generative retrieval. Across these settings, LLaMA models consistently outperform T5 models, suggesting a particular advantage for larger decoder-only models in generative retrieval. Our findings underscore that model sizes, data availability, and inference computation interact to unlock the full potential of generative retrieval, offering new insights for designing and optimizing future systems. We release code at \href{https://github.com/HongruCai/SLGR}{SLGR GitHub repository}.
\end{abstract}




\maketitle

\input{1_intro}

\input{2_related_work}

\input{3_method}
\input{4_scalinglaws}

\input{5_testtime}
\input{6_discussion}
\input{7_conclusion}

\begin{acks}
This research is supported by the National Research Foundation, Singapore under its National Large Language Models Funding Initiative (AISG Award No: AISG-NMLP-2024-002), Research Grants Council of Hong Kong(PolyU/15209724) and PolyU internal grants (BDWP). Any opinions, findings and conclusions or recommendations expressed in this material are those of the authors and do not reflect the views of National Research Foundation, Singapore.
\end{acks}

\bibliographystyle{ACM-Reference-Format}
\balance
\bibliography{sample-base}

\end{document}

%% file: 1_intro.tex
\section{Introduction}

Document retrieval is a fundamental area in information retrieval, focusing on retrieving relevant documents from large-scale corpora in response to user queries. Early retrieval systems were built on term-based heuristic methods, such as TF-IDF~\cite{salton1975tfidf} and BM25~\cite{rob2009bm25}, which rely on query and document term overlap. With the development of pre-trained language models, like BERT~\cite{devlin2019bert}, retrieval evolved into the dense retrieval paradigm, where queries and documents are mapped into a shared high-dimensional vector space, achieving advanced performance in document retrieval. Recently, with the rise of generative large language models (LLMs)~\cite{openai2024gpt4technicalreport,touvron2023llama2openfoundation,colin2020t5}, a new paradigm called generative retrieval has emerged~\cite{li2024surveygenerativesearchrecommendation}. Instead of \textbf{\textit{matching}} queries with documents, generative retrieval directly \textbf{\textit{generates}} documents based on a given query. By reformulating the retrieval task as an autoregressive generation problem, generative retrieval indeed introduces a novel solution to the research field.


\begin{figure*}[]
\setlength{\abovecaptionskip}{-0.00cm}
\centering
\includegraphics[width=0.95\linewidth]{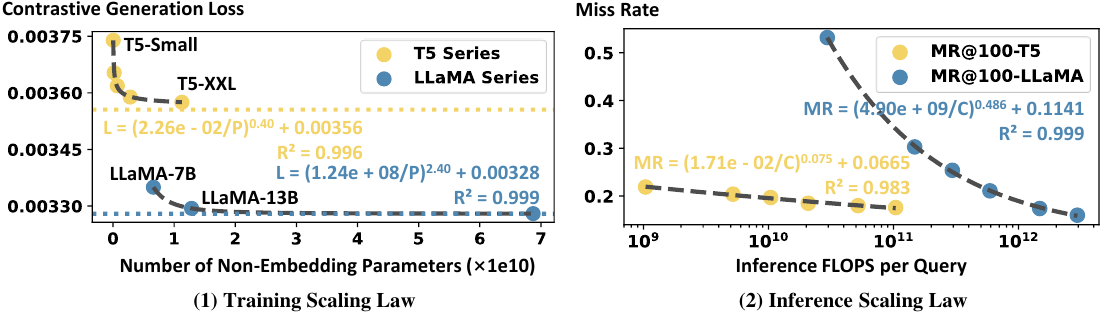}
\caption{Scaling laws of n-gram-based methods in training and inference. (1) Training Scaling Law: Contrastive Generation Loss shows a power-law relationship with model size for both T5 and LLaMA models. (2) Inference Scaling Law: Miss Rate decreases consistently with increasing inference FLOPs across the T5-Base and LLaMA-7B model.}
\vspace{-1.5em}
\label{Fig: sl_intro}
\end{figure*}

A central challenge in generative retrieval lies in designing effective document identifiers to represent documents, as generating entire long documents is impractical. The current identifiers can be divided into two broad categories based on how they carry semantics. 1) \textbf{Natural identifiers} retain inherent semantic information by leveraging components like titles~\cite{decao2021genre, lee2022gmhr, chen2022gere, li2023gcqa} or n-gram fragments~\cite{becil2022seal, chen2023aunified, li2023minder, wang2023novo} extracted from the original text. Titles provide concise, human-readable overviews, while n-gram snippets capture more granular semantic features. 2) \textbf{Learned identifiers}, on the other hand, derive semantic representations through clustering or codebook methods. Notable examples include numeric IDs~\cite{tay2022dsi, wang2022nci, zhuang2023nciqg, mehta2023dsipp} and codebook-derived tokens~\cite{yang2023asi, sun2023learntotokenize, zeng2024ripor, zhang2024mevi}, which discretize document embeddings into token sequences. Existing generative retrieval methods within these two categories have continued to evolve, showing promising performance.

Despite these advancements, the core advantages of generative retrieval remain unclear, with no established consensus in the research community. One key reason for success in many LLM-based tasks is scaling—increasing model size, data volume, and inference computation~\cite{brown2020llmfewshot, wei2022emergent, wu2024inferencescalinglawsempirical}. Given that generative retrieval follows the same autoregressive paradigm and is even built on LLM backbones, it is much more meaningful to explore scaling laws in generative retrieval to unlock the full potential of this paradigm.

While scaling laws have been extensively studied in various domains~\cite{zhai2022scaling,rad2023robustspeech,jia2021scalinupvl, rad2021clip,zhai2025hstu,zhang2024wukong}, exploring scaling in generative retrieval remains highly challenging. 1) To date, most studies have used relatively small encoder-decoder architectures (e.g., BART~\cite{lewis2019bart}, T5~\cite{colin2020t5}) rather than larger, modern LLMs like LLaMA~\cite{touvron2023llama2openfoundation}. 2) Moreover, recent breakthroughs in LLM scaling have centered on \emph{decoder-only} models—highly effective for generative tasks yet rarely explored for retrieval. 3) Standard retrieval metrics (e.g., recall, NDCG) are discrete and may miss nuanced performance variations, while contrastive entropy metrics from dense retrieval~\cite{fang2024sldr} are not suited for generative setups lacking a direct query-document scoring mechanism. 4) Additional complexity arises from the diverse ways of constructing document identifiers—whether natural or learned identifiers—each of which may respond differently to model scaling.

To address the above challenges, we introduce a new metric and employ larger models across different retrieval methods to systematically investigate how model size, training data scale, and inference compute impact performance. To capture subtle performance variations beyond discrete metrics, we propose a novel evaluation metric inspired by contrastive entropy and generation loss in neural scaling laws~\cite{kaplan2020scalinglawsneurallanguage}. Our metric measures the probability of generating the correct document identifier for a given query while considering random negative samples, yielding a continuous and sensitive performance signal. This metric enables consistent comparisons across various generative retrieval methods, models, and data scales. Leveraging this metric, we conduct extensive experiments to uncover the scaling behaviors of generative retrieval under different model sizes and data sizes. Additionally, we analyze how increased inference-time computation influences performance, highlighting its role in improving retrieval accuracy.


From our extensive exploration, several intriguing findings stand out. 1) We observe that n-gram-based generative retrieval aligns remarkably well with both training and inference scaling laws, as illustrated in Figure~\ref{Fig: sl_intro}, which presents clear scaling curves under varying model sizes and inference computation. 2) Expanding the training data scale benefits all methods, and n-gram-based approaches demonstrate especially robust gains, indicating a strong synergy between LLMs and natural identifiers. 3) We discover that LLaMA models consistently outperform T5 models and exhibit higher theoretical upper bounds, hinting that the generative ability of larger decoder-only models may be particularly advantageous for generative retrieval. 4) We find that boosting inference computation yields clear performance improvements that follow power-law trends, revealing that generative retrieval can significantly profit from additional inference computation—an aspect rarely discussed in prior work.

%% file: 2_related_work.tex
\section{Related Work}

In this section, we revisit the previous studies of generative retrieval and scaling laws. We first present the key advancements in generative retrieval, focusing on document identifier design and training methodologies. Then, we discuss neural scaling laws and their applications in various domains, including retrieval tasks.

\subsection{Generative Retrieval}

Generative retrieval reformulates retrieval as an autoregressive generation task, where a language model directly generates the identifier of the target document given a query. A central component in generative retrieval is the document identifier (DocID), which serves as the generation target. DocID can be broadly categorized into two types: natural identifiers and learned identifiers. 

The first type uses identifiers that naturally carry semantic meaning about its associated document, which are comparably cost-efficient to be created, without the need for additional human supervision or forcing any structure in the search space. A line of work~\cite{decao2021genre, lee2022gmhr, chen2022gere, li2023gcqa} explored titles as identifiers and DSI~\cite{tay2022dsi} tested the first N words of the passage, providing human-readable summaries of document content. Multiple textual fragments ~\cite{becil2022seal, chen2023aunified,li2023minder, wang2023novo} extracted directly from the document are then proved to capture richer semantic features of the document. These n-grams are not tied to a specific document but reflect shared semantic content across multiple semantically related documents. The second type called learned identifiers, acquires semantic meaning through dense representation of passages clustering or codebook training, which induces structure in the search space with semantically similar documents having more similar document IDs. In this case, each document is assigned a unique numerical representation, and retrieval is achieved by generating the corresponding identifier. Specifically, numeric ID~\cite{tay2022dsi, wang2022nci, zhuang2023nciqg, mehta2023dsipp} was found easy to construct but required extra memory steps. And codebook-derived tokens~\cite{yang2023asi, sun2023learntotokenize, zeng2024ripor, zhang2024mevi} have been proven effective because the search space is reduced after each decoding step.

Generative retrieval research has also advanced in training methods, which can be divided into generative training and discriminative training. In generative training, the model is trained to generate the appropriate identifier for a given query~\cite{pradeep2023generativescaleto,tay2022dsi, wang2022nci, zeng2024ripor, becil2022seal}, aligning naturally with the generative capabilities of LLMs to produce accurate document identifiers. Discriminative training, on the other hand, employs ranking losses~\cite{li2024learntorank, tang2024listwise} and negative sample mining techniques~\cite{zeng2024ripor} to teach the model to produce a ranked list of documents. These approaches align the training objectives with the ranking requirements of retrieval tasks.

\subsection{Neural Scaling Laws}

Neural scaling laws describe predictable patterns of performance improvement as model size, dataset size, and computational resources increase. Baidu~\cite{hestness2017deeplearningscalingpredictable} first introduced power-law relationships between test loss and these factors, offering an insight to predict neural network training. OpenAI~\cite{kaplan2020scalinglawsneurallanguage} extended this concept to larger models, demonstrating that model scaling yields consistent improvements in tasks like language modeling. Google~\cite{hoff2022anemperical} further introduced a unified formula for scaling laws, thus laying the groundwork for scaling strategies in neural networks.

Scaling laws have been successfully applied to various domain-specific fields such as speech recognition~\cite{rad2023robustspeech}, computer vision~\cite{deh2023scalingvision, zhai2022scaling}, and vision-language models~\cite{jia2021scalinupvl, rad2021clip}. In the field of information retrieval, scaling laws have been explored in recommendation~\cite{dealrec}. Studies have examined applications in Click-Through Rate (CTR) prediction~\cite{ardalani2022understandingscalinglawsrecommendation} and sequential recommendation models~\cite{zhang2024salinglargeseq} using unique item identifiers. Recent research has demonstrated the effectiveness of trillion-parameter sequential transducers for generative recommendations~\cite{zhai2025hstu} and the development of architectures like Wukong~\cite{zhang2024wukong} has established scaling laws for large-scale recommendation systems by effectively capturing diverse, high-order interactions through scalable network layers. 

However, research on scaling laws in retrieval tasks remains limited. Scaling laws for dense retrieval~\cite{fang2024sldr} have been investigated, focusing on embedding-based methods using BERT-like models~\cite{devlin2019bert}. Another study on industrial multi-stage advertisement retrieval systems~\cite{wang2024scalinglawsonlineadvertisement} emphasizes task-specific optimizations but lacks exploration of general scaling laws applicable to generative retrieval. An early exploration of generative retrieval examines performance across varying corpus and model sizes~\cite{pradeep2023generativescaleto} but is limited to learned identifier methods, leaving natural identifier methods unaddressed. Additionally, it only adopts different sizes of T5 models and does not fit power-law scaling relationships.

In contrast to prior work, our study investigates generative retrieval with larger decoder-only models, compares both learned and natural DocID strategies, and examines whether power-law relationships hold across model and data scales. Moreover, we analyze inference-time scaling behavior, which has received little attention in previous studies. These distinctions enable a more comprehensive understanding of generative retrieval scaling beyond existing task-specific or partial analyses.

%% file: 3_method.tex
\section{Methodology}

In this section, we first formalize the generative retrieval task in Section 3.1. Next, we describe the representative approaches for natural identifier and learned identifier: MINDER~\cite{li2023minder} and RIPOR~\cite{zeng2024ripor}, respectively, in Section 3.2. The backbones and training configurations used for generative retrieval are introduced in Section 3.3. Finally, to address the limitations of traditional ranking metrics in reflecting the scaling behaviors of generative retrieval, we propose contrastive generation loss in Section 3.4.

\subsection{Problem Formulation}

We formalize the generative retrieval task as a two-step process. 1) Identifier assignment: assigning identifiers to documents and 2) Identifier Generation: generating query-specific identifiers to retrieve relevant documents.

\textbf{Identifier Assignment.} Let a corpus of documents be denoted as \( \mathcal{D} = \{d_1, d_2, \ldots, d_N\} \). Each document \( d \in \mathcal{D} \) is associated with a set of identifiers \( \mathcal{I}_d = \{i_d^1, i_d^2, \ldots, i_d^M\} \) through a transformation function \( h(d; \psi) \), parameterized by \( \psi \):  

\begin{equation}
\mathcal{I}_d = h(d; \psi).
\end{equation}

\textbf{Identifier Generation.}
During training, the generative language models $f$ parameterized by \( \theta \), learn to generate query-specific identifiers \( \mathcal{I}_q = \{i_q^1, i_q^2, \ldots, i_q^K\} \) based on paired training data consisting of queries and relevant document identifiers. During inference, given a query \( q \), the model $f$ use $q$ as input and generate a set of identifiers:  

\begin{equation}
   \mathcal{I}_q = f(q; \theta). 
\end{equation}
The generated identifiers \( \mathcal{I}_q \) are then used to retrieve documents \( \mathcal{D}_q \) by applying the reverse of the identifier assignment function. 

Different generative approaches usually focus on designing different types of identifiers. However, the goal remains the same: to optimize the transformation function \( h(d; \psi) \) and the generative language model \( f(q; \theta) \) to generate query-specific identifiers that maximize the relevance of the retrieved documents \( \mathcal{D}_q \).

\subsection{Representative Generative Retrieval Methods}

As outlined above, the document identifier is a crucial component of generative retrieval. There are two primary categories: natural identifiers and learned identifiers. In this study, we select representative techniques from each category to evaluate their scaling behaviors. For \textbf{natural identifiers}, we use the n-gram-based method as n-grams effectively capture diverse semantic relationships between queries and documents. For \textbf{learned identifiers}, we adopt a codebook-based approach, as it leverages advanced neural methods to encode semantic information effectively.

\subsubsection{N-gram-based Generative Retrieval}
\label{sec:n-gram}

N-grams offer a flexible and query-adaptive approach to document identifiers~\cite{becil2022seal, chen2023aunified, li2023minder, wang2023novo}. These identifiers are directly extracted from documents based on their overlap or semantic relevance to specific queries, capturing key contextual features that align closely with user queries.

In this method, an LLM is trained to generate query-specific n-grams, which act as identifiers for ranking documents. As shown in Figure~\ref{Fig: docid} (1), the training process begins by selecting n-grams from documents based on their overlap or semantic similarity with user queries. These n-grams serve as the basis for training the LLM to predict relevant identifiers given a query. At inference, the LLM generates n-grams for a query, and these are used to score and rank documents based on a heuristic function, such as n-gram frequency or semantic similarity. For our experiments, we adopt the method of MINDER~\cite{li2023minder}, focusing on extracting identifiers from the body text of documents.

\subsubsection{Codebook-Based Generative Retrieval}

Codebooks originate from techniques designed to create discrete visual data representations~\cite{lee2022rqvae, van2017neuraldiscrete}. Learned codebooks for documents represent them as sequences of unique codes that effectively capture the semantics of their associated content~\cite{yang2023asi, sun2023learntotokenize, zeng2024ripor, zhang2024mevi}. 

\begin{figure}[]
\setlength{\abovecaptionskip}{-0.00cm}
\centering
\includegraphics[width=\linewidth]{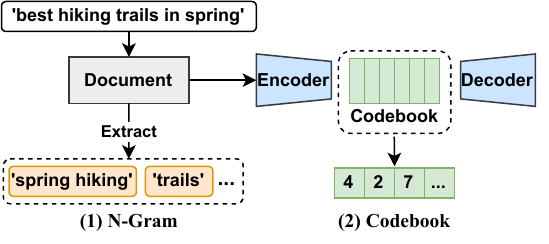}
\caption{Comparison of n-gram-based and codebook-based methods. N-grams are extracted text spans, while codebook-based methods are discrete representations generated via codebook.}
\label{Fig: docid}
\vspace{-2em}
\end{figure}

As shown in Figure~\ref{Fig: docid} (2), the process begins by encoding documents into dense vector representations using an encoder network. These dense vectors are then discretized into tokens by mapping them to entries in the learned codebook. Finally, a decoder network reconstructs the original document from the codebook to ensure the generated representations are accurate and compact. The resulting code sequences serve as unique identifiers for documents, establishing a one-to-one correspondence between the code sequence and the document. At inference, the LLM generates a code sequence for a query, which is matched to the corresponding document based on the learned codebook. In our study, we select the RIPOR~\cite{zeng2024ripor} as the representative of the codebook-based method.

\subsection{Backbone Models and Training Setting}

To investigate the scaling capabilities of different generative retrieval systems, it is essential to first identify the backbone models they utilize and their corresponding training settings.

\subsubsection{Backbone Models}

For our experiments, we use the widely adopted T5~\cite{colin2020t5} series as the primary backbone, which has been extensively employed in previous generative retrieval studies~\cite{tay2022dsi,zeng2024ripor,ren2023tome,mehta2023dsipp,zhuang2023dsiqg,zhang2024mevi}. To evaluate the effect of model size, we experiment with all T5 variants: T5-Small, T5-Base, T5-Large, T5-XL, and T5-XXL, which differ only in parameter sizes while maintaining identical pre-training configurations. 

To further investigate architectures beyond encoder-decoder models and explore the impact of larger parameter scales, we also experiment with LLaMA models. Specifically, we employ three sizes of LLaMA-2~\cite{touvron2023llama2openfoundation} models: LLaMA-2-7B, LLaMA-2-13B, and LLaMA-2-70B. These decoder-only models not only represent a different architectural paradigm but also significantly increase the parameter scale compared to typical encoder-decoder backbones. Their scaling properties observed in other tasks provide valuable insights into how architecture and model size influence generative retrieval performance.

\subsubsection{Training Setting}

The training process involves two distinct setups corresponding to the two generative retrieval approaches: n-gram-based generative retrieval and codebook-based generative retrieval. In the following sections, we detail the datasets, training configurations, and loss functions for the two representative approaches, respectively.

\vspace{3pt}
\noindent$\bullet$ \textbf{N-Gram-Based Generative Retrieval.} 
For n-gram-based generative retrieval, we use the Natural Questions (NQ)~\cite{kwiatkowski2019nq} dataset, which contains over 20 million documents. Following the MINDER methodology, which originally involves three types of identifiers, we focus exclusively on the body text as the identifier type. This simplification is made because our study emphasizes scaling behavior rather than achieving absolute performance, and the body text serves as the most important identifier type. For each document, we select 10 n-grams, each consisting of 10 tokens, based on their overlap with the associated query to ensure semantic relevance. The final training set includes nearly 600,000 query-to-n-gram pairs.

During training, the input consists of the query, and the label is a single n-gram from the corresponding document. The training objective is to minimize the cross-entropy loss for generating each n-gram \( n \), given the query \( q \):  

\begin{equation} 
\mathcal{L}_{\text{n-gram}} = - \log P(n \mid q; \theta),
\end{equation}  
where \( n \) represents an individual n-gram and $\theta$ represents the model parameters. Since each n-gram is treated independently, the model learns to predict each query-relevant n-gram as a separate target.

To keep consistent with the following setup of codebook-based methods, we train each model for one epoch using the MINDER-provided data with LoRA~\cite{hu2022lora}. As recommended in MINDER, the learning rate for T5 models is set to 3e-5. For LLaMA models, we use a learning rate of 3e-4, a commonly adopted value. 

\vspace{3pt}
\noindent$\bullet$ \textbf{Codebook-Based Generative Retrieval.} 
Following the selected representative codebook-based generative retrieval method, RIPOR~\cite{zeng2024ripor}, we conduct experiments using the MSMARCO-1M dataset, a subset of the MSMARCO~\cite{nguyen2017msmarco} dataset, containing one million passages and query-document pairs. This dataset is chosen because RIPOR is only available on MSMARCO, and since our study focuses on scaling behavior rather than absolute performance, the choice of the dataset has minimal impact on the relative results. 

The codebook consists of \( N_c \) unique codes, with each document represented as a sequence of \( L_c \) codes. These codes are treated as new tokens added to the vocabulary of the LLM $\theta$. The training follows a standard generative setup, where the input is the query and the output is the document's code sequence. The training objective is to minimize the cross-entropy loss, which measures the negative log-likelihood of generating the correct code sequence \( \mathbf{c}_d = \{c_1, c_2, \ldots, c_{L_c}\} \) for document \( d \), given query \( q \):  

\begin{equation}
   \mathcal{L}_{\text{codebook}} = - \sum_{t=1}^{L_c} \log P(c_t \mid q, c_{<t}; \theta). 
\end{equation}

In our experiments, \( N_c = 256 \) unique codes, and \( L_c = 32 \) codes per document. We train each model for one epoch using the RIPOR-provided data with LoRA~\cite{hu2022lora}. For T5 models, the learning rate is set to 1e-3, consistent with configurations in RIPOR. For LLaMA models, the learning rate is set to 3e-4.

\subsection{Evaluation}

Evaluating generative retrieval models requires metrics that effectively capture nuanced variations of retrieval performance. Traditional retrieval metrics, such as NDCG and Recall, are not well-suited for this purpose. First, these metrics are inherently discrete, making them incapable of capturing fine-grained differences in model outputs. Second, they primarily evaluate changes in ranked lists, offering limited insight into the nuanced behavior of model outputs. Lastly, their reliance on cutoff parameters means they only consider documents within a specific range (e.g., top $K$), ignoring contributions from documents ranked lower, which limits their effectiveness for studying scaling behaviors in generative retrieval.

To address these limitations, we draw inspiration from prior work on dense retrieval metrics~\cite{fang2024sldr} and scaling laws in large language models~\cite{kaplan2020scalinglawsneurallanguage} to propose a novel evaluation metric tailored for generative retrieval. Building on the concept of contrastive entropy used in dense retrieval, we adapt it to the generative setting by incorporating the loss associated with query-to-identifier generation.

\vspace{3pt}
\noindent$\bullet$ \textbf{Contrastive generation loss.} 
In generative retrieval, the primary objective is to generate identifiers that correspond to relevant documents while distinguishing them from irrelevant ones. To quantify this ability, we define a contrastive generation loss (CGL), which evaluates the model's capacity to generate identifiers for positive documents in the presence of negative documents. 

For a query \( q \) and its associated positive document \( d^+ \), let \( \mathcal{I}_{d^+} \) be the identifier for \( d^+ \), and let \( \mathcal{I}_{d^-} \) be the identifiers for a set of negative documents \( \mathcal{D}^- \). The contrastive generation loss \(\mathcal{L_{\text{CGL}}}\) is defined as:  

\begin{equation}
\mathcal{L}_{\text{CGL}} = - \log \frac{\sum_{d^-} \mathcal{L}(q, \mathcal{I}_{d^-})}{\mathcal{L}(q, \mathcal{I}_{d^+}) + \sum_{d^-} \mathcal{L}(q, \mathcal{I}_{d^-})},
\end{equation}  
where \( \mathcal{L}(q, \mathcal{I}) \) represents the generation loss for the query \( q \) to produce the identifier \( \mathcal{I} \), calculated using the cross-entropy loss. 

Unlike contrastive loss~\cite{khosla2020supervised} or contrastive entropy~\cite{fang2024sldr} which operates in embedding space, CGL directly uses generation loss to assess the model’s preference for positive identifiers. It requires no additional training objective and works purely in the generation setting. CGL has the following advantages:

\setlength{\leftmargini}{10pt} \begin{itemize}

\item \textbf{Method compatibility.}
CGL supports both n-gram-based and codebook-based generative retrieval. For n-gram identifiers, it averages the generation loss overall n-grams of the positive document. For codebook methods, it evaluates the loss of generating the complete token sequence assigned to each document.

\item \textbf{Model compatibility.}
CGL is applicable across different LLMs, as it operates solely on forward generation loss without requiring model-specific structures or additional training objectives.

\item \textbf{Relative evaluation.}
By using the ratio between positive and negative losses, CGL mitigates sensitivity to absolute loss values, ensuring consistent evaluation.
\end{itemize}

\begin{figure}[]
\setlength{\abovecaptionskip}{-0.00cm}
\centering
\includegraphics[width=\linewidth]{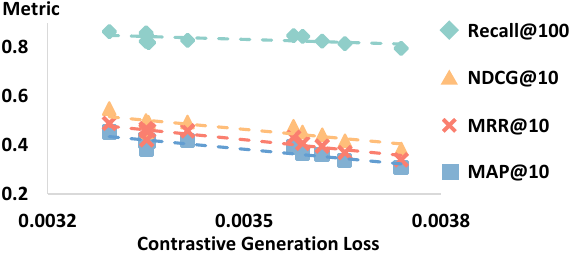}
\caption{Relationship between proposed Contrastive Generation Loss and traditional retrieval metrics (Recall@100, NDCG@10, MRR@10, and MAP@10). The results demonstrate a nearly linear correlation, validating the effectiveness of CGL in reflecting retrieval performance as measured by traditional metrics.}
\label{Fig: cor_metric}
\vspace{-2em}
\end{figure}

\vspace{3pt}
\noindent$\bullet$ \textbf{Validation.} 
To validate the proposed CGL, we conducted experiments to analyze its relationship with existing ranking metrics, such as Recall, NDCG, MAP, and MRR. Using the MINDER framework as an example, we trained models of varying sizes and configurations and then calculated their retrieval performance using Recall@100, NDCG@10, MAP@10, MRR@10 as well as the proposed CGL. Results in Figure~\ref{Fig: cor_metric} reveal an almost linear relationship between contrastive generation loss and standard metrics. This alignment demonstrates that the proposed metric effectively captures retrieval performance, offering a consistent and reliable evaluation framework across different settings.

%% file: 4_scalinglaws.tex
\begin{figure*}[]
\setlength{\abovecaptionskip}{-0.02cm}
\setlength{\belowcaptionskip}{-0.10cm}
\centering
\includegraphics[width=0.95\linewidth]{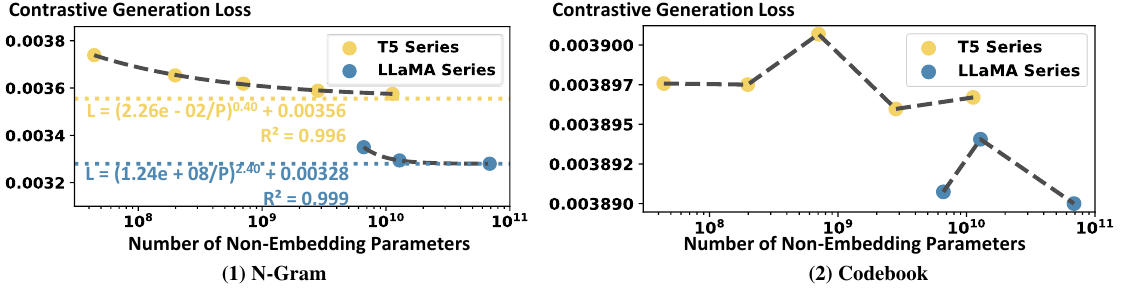}
\caption{Scaling behavior of contrastive generation loss regarding model size for (1) n-gram-based and (2) codebook-based method. The results demonstrate a clear scaling trend for the n-gram-based approach, while the codebook-based method exhibits no consistent improvement as model size increases.}
\vspace{-1em}
\label{Fig: scaling_model}
\end{figure*}

\section{Training Scaling Laws}

In this section, we present the results of our experiments and summarize our investigation into the training scaling laws for generative retrieval. Specifically, we analyze how model size, training data size, and identifier methods influence retrieval performance, using CGL as the evaluation metric. 

\subsection{Model Size Scaling}

We now investigate the impact of model size on retrieval performance. We begin by examining n-gram-based generative retrieval.

\subsubsection{N-Gram-Based Generative Retrieval}

We investigated the effect of model size on generative retrieval performance using the n-gram-based method. Models of varying sizes, including T5 and LLaMA, were fine-tuned on query-to-n-gram training pairs, and their performance was evaluated using the CGL on the test set.  Figure~\ref{Fig: scaling_model} (1) illustrates the scaling behavior of T5 and LLaMA models concerning this metric. Based on the observed relationship between model size and the CGL, we propose a scaling law to quantify this behavior as follows:  

\begin{equation}
\mathcal{L}_{\text{CGL}}(P) = \left(\frac{\gamma}{P}\right)^{\alpha} + \lambda_P.
\end{equation}
Here, $P$ represents the number of non-embedding parameters of the model, and $\mathcal{L}_{\text{CGL}}(P)$ denotes the CGL on the test set. The parameters $\gamma$, $\alpha$, and $\lambda_P$ are coefficients determined through fitting. Here, $\lambda_P$ represents the irreducible loss, a theoretical lower bound on performance as $P$ approaches infinity, accounting for limitations such as dataset noise and variability in relevance judgments.

Using the least squares method, we fit the scaling law and report the fitted coefficients for T5 and LLaMA in Table~\ref{tab:scaling_param_minder}. The results reveal several important insights: 1) Both models demonstrate a strong power-law relationship between model size and contrastive generation loss, with exceptionally high coefficients of determination. 2) LLaMA demonstrates comprehensive performance advantages over T5, characterized by a more efficient scaling mechanism. Specifically, LLaMA achieves lower CGL across model sizes and exhibits a steeper improvement curve (scaling exponent of $\alpha = 2.40$ versus T5's $\alpha = 0.40$). LLaMA's smaller irreducible loss ($\lambda_P$) suggests a higher potential performance ceiling, indicating its superior capability to approach the theoretical limits of generative retrieval performance. These findings highlight LLaMA's promising performance and signal the potential of decoder-only architectures.

\begin{table}[]
\centering
\caption{Fitted parameters for the scaling law on model sizes with n-gram-based methods.}
\vspace{-1em}
\begin{tabular}{ccccccc}
\hline
Method & Model & $\gamma$ & $\alpha$ & $\lambda_P$ & $R^2$ \\
\hline
N-Gram             & T5 Series           & $2.26 \times 10^{-2}$ & $0.40$ & $0.00356$ & $0.996$ \\
N-Gram             & LLaMA Series       & $1.24 \times 10^{8}$  & $2.40$ & $0.00328$ & $0.999$ \\
\hline
\end{tabular}
\vspace{-2em}
\label{tab:scaling_param_minder}
\end{table}

\subsubsection{Codebook-Based Generative Retrieval} 

For the codebook-based method, we conducted similar experiments, fine-tuning different sizes of T5 and LLaMA models on query-to-code sequence training pairs and evaluating their performance using contrastive generation loss on the test set. The results, shown in Figure~\ref{Fig: scaling_model} (2), reveal that neither T5 nor LLaMA models exhibit a consistent reduction in CGL as the model size increases, with the value fluctuating across different model sizes and showing no clear scaling trend. This suggests that increasing model size does not inherently enhance retrieval performance for codebook-based methods. This finding aligns with prior research~\cite{pradeep2023generativescaleto}, where T5-XXL underperformed smaller T5-XL with similar generative retrieval methods. 

The possible reasons are as follows: 1) Codebook tokens are newly introduced and unrelated to the models' pretraining objectives, requiring the models to learn entirely new semantic relationships during fine-tuning. 2) Newly introduced tokens often demand more extensive training to be fully integrated into the model's generative capabilities. In our experiments, only a single epoch of fine-tuning was conducted, which may not have been sufficient for the models to fully learn the codebook representations. Scaling behavior might emerge with additional training epochs once the models better understand these novel tokens. We leave this possibility for future research as the computational intensity needed makes comprehensive exploration impractical for us.

Despite the lack of scaling trends, LLaMA consistently achieves lower CGL than T5 across all model sizes, highlighting its stronger retrieval capabilities. These findings suggest LLaMA demonstrates promising performance characteristics and may signal avenues for future research in generative retrieval.

\begin{figure*}[]
\setlength{\abovecaptionskip}{-0.05cm}
\centering
\includegraphics[width=0.95\linewidth]{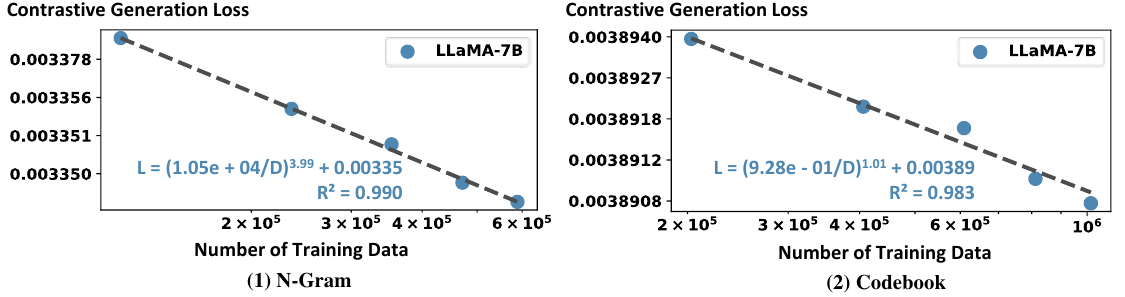}
\caption{Scaling behavior of contrastive generation loss concerning training data size for n-gram-based (left) and codebook-based (right) methods using the LLaMA-7B model. The results show clear scaling trends for both methods, with a steeper decline observed for n-grams-based generative retrieval.}
\vspace{-1em}
\label{Fig: scaling_data}
\end{figure*}

\subsubsection{Comparsions.} 
The results presented in Figure~\ref{Fig: scaling_model} highlight clear differences between the n-gram-based and codebook-based methods in terms of scaling behavior and overall performance. 1) The n-gram-based method significantly outperforms the codebook-based approach, demonstrating lower CGL. Even the most advanced LLaMA models utilizing codebook tokens cannot match the performance of T5 models with n-gram-based retrieval. This performance gap highlights the intrinsic challenges of codebook tokens, which lack the semantic coherence and natural language alignment inherent in n-grams. 2) LLaMA consistently outperforms T5 across both methods, achieving lower CGL at comparable model sizes. This highlights LLaMA’s stronger generative capabilities and its architectural advantage.

\subsection{Data Size Scaling}

The size of the training dataset also plays a critical role in determining the performance of generative retrieval models. In this section, we investigate how varying the training data size influences retrieval performance while keeping other factors, such as model size and architecture, constant.

\subsubsection{N-Gram-Based Generative Retrieval}

To study the effect of training data size, we use the $ \text{LLaMA-2-7B} $ model and incrementally increase the number of training pairs constructed using the n-gram-based method. Figure~\ref{Fig: scaling_data} shows the scaling behavior of the contrastive generation loss ($ \mathcal{L}_{\text{CGL}} $) concerning training data size. Similar to model size, we fit the scaling behavior using the following power-law equation:

\begin{equation}
\mathcal{L}_{\text{CGL}}(D) = \left(\frac{\eta}{D}\right)^{\beta} + \lambda_D  
\end{equation}
Here, $ D $ represents the number of query-identifier pairs, and $ \mathcal{L}_{\text{CGL}}(D)$ denotes the CGL on the test set. The parameters $ \eta $, $ \beta $, and $ \lambda_D $ are coefficients determined through fitting. The term $ \lambda_D $ represents the irreducible loss, a theoretical lower bound on retrieval performance as $ D $ approaches infinity. 

Using the least squares method, we fit the scaling law to the observed data, achieving a coefficient of determination of $ R^2 = 0.990 $, which indicates a strong fit. As seen in Figure~\ref{Fig: scaling_data} (1), retrieval performance improves significantly as the training data size increases, with the CGL decreasing sharply. The power-law scaling behavior reflects the model's capacity to leverage larger datasets to better capture the semantic relationships between queries and identifiers. 

\begin{table}[]
\centering
\caption{Fitted parameters for the scaling law on data sizes.}
\vspace{-0.5em}
\begin{tabular}{ccccccc}
\hline
Method & Model & $\eta$ & $\beta$ & $\lambda_D$ & $R^2$ \\
\hline
N-Gram             & LLaMA-7B         & $1.05 \times 10^{4}$  & $3.99$ & $0.00335$ & $0.990$ \\
Codebook           & LLaMA-7B         & $9.28 \times 10^{-1}$ & $1.01$ & $0.00389$ & $0.983$ \\
\hline
\end{tabular}
\vspace{-1,5em}
\label{tab:scaling_data}
\end{table}

\subsubsection{Codebook-Based Generative Retrieval} 

We also evaluated the effect of training data size on retrieval performance for the codebook-based method. Similar to the n-gram-based experiment, the $ \text{LLaMA-2-7B} $ model was fine-tuned on training datasets of varying sizes, and constructed with query-code sequence pairs. 

Using the same power-law equation as for n-grams, we fit the scaling behavior of the codebook-based method. The fitted curve in Figure~\ref{Fig: scaling_data} (2) achieves a coefficient of determination of $ R^2 = 0.983 $, indicating a strong fit. As the training data size increases, the CGL decreases steadily, demonstrating that retrieval performance improves with larger datasets. The results highlight that even for the codebook-based method, which involves learning entirely new representations unrelated to the model's pretraining objectives, increasing the data size leads to performance enhancements. 

\subsubsection{Comparsions.} 

The results in Figure~\ref{Fig: scaling_data} highlight key differences between n-gram-based and codebook-based methods in their scaling behavior and overall retrieval performance.

For n-gram-based methods, the scaling exponent (\( \beta = 3.99 \)) is much larger than that of codebook-based methods (\( \beta = 1.01 \)), indicating a steeper improvement in performance with increased data size. This can be attributed to the semantic richness of n-grams, which align closely with the model's pretraining objectives, allowing the model to fully leverage larger datasets. In contrast, the codebook-based method lacks such alignment, resulting in a slower rate of improvement as data size increases. The low scaling exponent implies that this method requires substantially more training data to achieve comparable performance. Recent studies suggest more advanced training strategies, such as ranking losses~\cite{tang2024listwise, zeng2024ripor}, could potentially address these learning challenges.

\subsection{Model-Data Joint Laws}

To capture the joint effects of model size and data size on retrieval performance, we combine the observations from the previous sections into a single scaling function. Inspired by established scaling laws in LLMs~\cite{kaplan2020scalinglawsneurallanguage}, we employ the following equation to describe the combined effects:

\begin{equation}
\mathcal{L}(P, D) = \left( \left(\frac{\gamma}{P}\right)^{\frac{\alpha}{\beta}} + \frac{\eta}{D} \right)^{\beta} + \delta.
\end{equation}
Here, \( P \) and \( D \) represent the model size (number of non-embedding parameters) and training data size, respectively. The parameters \( \gamma \), \( \eta \), \( \alpha \), \( \beta \), and \( \delta \) are coefficients determined through fitting. Based on experimental results using LLaMA with n-gram-based method across various model sizes and training data sizes, we obtained the following estimates for these coefficients:

\begin{equation}
\gamma = 6.32 \times 10^3, \quad 
\alpha = 3.27, \quad 
\beta = 0.95,
\end{equation}
\begin{equation}
    \eta = 3.37 \times 10^5, \quad 
\delta = 3.26 \times 10^{-3}, \quad 
R^2=0.976.
\end{equation}

The coefficient of determination indicates a high degree of accuracy in capturing the relationship between model size, data size, and retrieval performance. This unified scaling function highlights the complementary contributions of model size and data size to retrieval performance. Larger models reduce loss by better capturing semantic relationships, while increased data size allows for improved learning of these relationships. The joint law provides a valuable framework for balancing model size and data requirements to optimize performance efficiently.

%% file: 5_testtime.tex
\section{Inference Scaling Laws}

In the previous sections, we demonstrated the existence of training scaling laws in generative retrieval concerning model size and data size. Beyond training, recent studies have also revealed another dimension of scaling: the scaling of computational investment during inference~\cite{wu2024inferencescalinglawsempirical}. Specifically, increasing inference computing, such as scaling the number of decoding tokens, yielded substantial performance gains. This raises the question of whether similar benefits extend to generative retrieval.

Inference scaling is particularly promising in generative retrieval because the identifiers used for retrieval are generated dynamically at this stage. The generation process, controlled by parameters such as beam size, directly affects the quality and diversity of identifiers, and hence retrieval performance. Motivated by these observations, we investigate whether inference scaling laws observed in LLMs also hold in generative retrieval, and how factors like beam size and compute budget impact its effectiveness.

\begin{figure}[]
\setlength{\belowcaptionskip}{-0.20cm}
\centering
\includegraphics[width=0.9\linewidth]{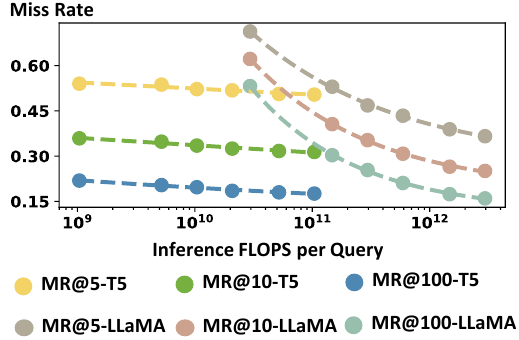}
\caption{Inference scaling behavior of n-gram-based methods across T5 and LLaMA models. Miss Rate consistently decreases as inference FLOPs per query increase, demonstrating a power-law relationship. LLaMA models show a steeper decline, particularly at higher inference FLOPs, highlighting their superior scalability compared to T5.}
\label{Fig: scaling_inf}
\vspace{-1em}
\end{figure}

\subsection{Experimental Setup}

To explore this, we focus on n-gram-based generative retrieval as a representative approach. N-gram-based methods are particularly suitable for investigating inference scaling because their retrieval process aligns well with the core principles of scaling during inference. Specifically, these methods first generate n-grams and then use these n-grams to score and rank documents. By increasing the beam size during n-gram generation, we can systematically expand the set of candidate n-grams. This increase not only boosts the quantity of n-grams available for scoring but also enhances their diversity and quality, which is likely to improve the final document retrieval performance.

\vspace{3pt}
\noindent$\bullet$ \textbf{Implementation details.} 
We use the n-gram-based generative retrieval method described in Section~\ref{sec:n-gram}. And we chose T5-Base and LLaMA-7B as representative models, both are fine-tuned under identical settings on the NQ dataset~\cite{kwiatkowski2019nq}, using the same query-to-n-gram pairs. Each model is trained for one epoch to ensure consistency across experiments.  

During inference, we vary the beam size, testing values of \( B = \{1, 5, 10, 20, 50, 100\} \) and record the corresponding inference computation needed. Beam size determines the number and search space of candidate n-grams generated per query, affecting the diversity of identifiers used for document retrieval. Increasing the beam size effectively increases the computational cost during inference, as a larger search space requires more floating-point operations (FLOPs) to generate candidate n-grams. 

\vspace{3pt}
\noindent$\bullet$ \textbf{Evaluation.} 
To evaluate retrieval performance, we define a metric called Miss Rate (MR), which measures the proportion of relevant documents that are not retrieved within the top \(k\) results. 
\begin{equation}
    \text{MR}@k = 1 - \text{Recall}@k,
\end{equation}
where \(k\) represents the number of retrieved documents considered (e.g., \(k = 5, 20, 100\)). MR provides a straightforward and interpretable view of retrieval effectiveness by focusing on the proportion of relevant documents missed. By analyzing MR@5, MR@20, and MR@100, we systematically evaluate how varying beam sizes influence retrieval performance across different levels of precision.

\begin{table}[]
\centering
\caption{Fitted parameters for the inference scaling law with n-gram-based methods.}
\vspace{-1em}
\begin{tabular}{ccccccc}
\hline
Model & Miss Rate & $\mu$ & $\sigma$ & $\lambda_C$ & $R^2$ \\
\hline
T5-Base             & MR@5           & $1.60 \times 10^{-4}$ & $0.0620$ & $0.3834$ & $0.915$ \\
T5-Base              & MR@20       & $3.85 \times 10^{-4}$  & $0.0508$ & $0.1276$ & $0.970$ \\
T5-Base              & MR@100       & $1.71 \times 10^{-2}$  & $0.0755$ & $0.0665$ & $0.983$ \\
LLaMA-7B             & MR@5       & $2.71 \times 10^{9}$  & $0.3479$ & $0.2779$ & $0.999$ \\
LLaMA-7B               & MR@20       & $4.16 \times 10^{9}$  & $0.4233$ & $0.1859$ & $0.999$ \\
LLaMA-7B               & MR@100       & $4.90 \times 10^{9}$  & $0.4862$ & $0.1141$ & $0.999$ \\
\hline
\end{tabular}
\vspace{-1.5em}
\label{tab:fitted_param_inf}
\end{table}

\subsection{Results}

The results of our experiments are summarized in Figure~\ref{Fig: scaling_inf}, which illustrates the MR across different inference FLOPs per query for both T5-Base and LLaMA-7B models. We evaluate MR at different retrieval thresholds (\(k = 5, 20, 100\)) to assess the retrieval performance under varying levels of precision.

To analyze the relationship between inference computational cost and retrieval performance, we propose a fitting function:

\begin{equation}
    \text{MR}(C) = \left( \frac{\mu}{C} \right)^\sigma + \lambda_C,
\end{equation}
where \(C\) represents the inference FLOPs per query, \(\mu\), \(\sigma\), and \(\lambda_C\) are parameters to fit, and \(\lambda_C\) is irreducible loss, a theoretical lower bound on retrieval performance as $ C $ approaches infinity. 

The fitted curves in Figure~\ref{Fig: scaling_inf}, reveal a consistent trend for both T5-Base and LLaMA-7B models: the MR decreases steadily as the inference FLOPs per query increase. This validates the effectiveness of inference scaling in improving generative retrieval performance. As shown in Table~\ref{tab:fitted_param_inf}, the proposed scaling law fits well with the experimental results. The results reveal several key findings:
\setlength{\leftmargini}{10pt} \begin{itemize}

\item \textbf{MR@k sensitivity.} The rate of MR decline varies with the retrieval threshold $k$. MR@5 decreases the fastest for both models, showing strong benefits from inference scaling in high-precision retrieval, while MR@100 declines more slowly, indicating reduced sensitivity in broader recall settings.

\item \textbf{Model performance under low compute.} When inference FLOPs are below \(10^{11}\), T5-Base achieves lower MR across all \(k\) than LLaMA-7B, suggesting that T5-Base is more efficient in compute-constrained scenarios.

\item \textbf{Model performance under high compute.} As FLOPs increase beyond \(10^{11}\), LLaMA-7B gradually surpasses T5-Base, achieving lower MR@100 and MR@20, demonstrating greater benefits from increased inference computation.

\item \textbf{Irreducible loss.} LLaMA-7B exhibits lower irreducible loss (\( \lambda_C \)) for MR@5, suggesting superior capacity for high-precision retrieval. In contrast, it shows a slightly higher irreducible loss for MR@20 and MR@100, indicating that the T5-Base retains an advantage in broader recall scenarios.

\end{itemize}

%% file: 6_discussion.tex
\section{Discussion}

In this work, we explored the training and inference scaling laws in generative retrieval, proposing metrics to analyze retrieval performance under varying conditions. While our findings offer useful insights, several limitations and open questions remain.

\vspace{3pt}
\noindent$\bullet$ \textbf{Limitation of CGL.} 
While the proposed Contrastive Generation Loss reflects the model preference for correct identifiers under consistent settings, it does not directly correlate with traditional metrics like Recall. For example, models with similar CGL values may have different recall, and a lower CGL does not guarantee better performance. CGL is effective for relative comparisons, but its interpretability as an absolute performance measure is limited.

\vspace{3pt}
\noindent$\bullet$ \textbf{Simplified training settings.} 
For simplicity, we adopt the standard generative cross-entropy loss without incorporating more advanced objectives. While this may be sufficient for n-gram-based methods, which match the model’s pretraining objective, it poses challenges for codebook-based methods that need to learn entirely new tokens or relationships, demanding more training data and epochs. Prior work has explored more effective alternatives, such as learning-to-rank (LTR) losses~\cite{li2024learntorank, tang2024listwise}, which may better capture the retrieval objective. Future research could revisit this direction with more advanced training objectives.

\vspace{3pt}
\noindent$\bullet$ \textbf{Model scaling of the codebook-based method.} 
In our experiments, codebook-based methods showed clear scaling with training data size but not with model size. This may be due to the greater learning difficulty of codebook representations. The training data and time in our experiments may have been insufficient, preventing the scaling effect from emerging. Previous studies have shown that model performance can undergo substantial improvement at a certain point~\cite{du2024understanding}. Although we attempted to use RIPOR’s full dataset (over 80M query-code pairs), the computational cost for LLaMA was unaffordable. Future work could revisit this setting to better understand the scaling potential of codebook-based methods.

\vspace{3pt}
\noindent$\bullet$ \textbf{Inference scaling of the codebook-based method.} 
Our inference scaling experiments focused on n-gram-based retrieval, showing improved performance with increased beam size and alignment with power-law scaling. Codebook methods may also benefit from inference-time strategies such as beam search~\cite{zeng2024pag}, but we leave this exploration to future work due to space and resource constraints.

\vspace{3pt}
\noindent$\bullet$ \textbf{Dataset inconsistency across methods.}
In our experiments, the n-gram-based and codebook-based methods were trained and evaluated on different datasets, as we directly adopted identifier sets from prior work to simplify implementation and isolate scaling effects. Consequently, the comparison between methods may be affected by dataset differences. A more controlled comparison would involve evaluating both methods on a unified dataset—a direction we leave for future work.

%% file: 7_conclusion.tex
\section{Conclusion and Future Work}


In this paper, we explored the scaling laws of generative retrieval across training and inference dimensions. We analyzed how model size, training data size, and retrieval methods influence performance. Using Contrastive Generation Loss as the metric, we found a clear power-law relationship between model size and retrieval performance for n-gram-based methods. Additionally, we observed consistent scaling trends with increasing data size for both n-gram-based and codebook-based methods. A comparison between architectures showed that LLaMA consistently outperformed T5 under identical experimental conditions, highlighting LLaMA's superior capability for advancing generative retrieval. Beyond training scaling, we also investigated inference scaling for n-gram methods and found that increasing inference-time computation followed a power-law trend, offering a complementary axis for performance gains. Collectively, our findings validate the effectiveness of systematically scaling model capacity, training data, and inference computation, suggesting pathways to further enhance generative retrieval methods.

While our study offers a foundation for understanding the scaling laws of generative retrieval, several directions remain for future exploration. First, adopting more effective training objectives, such as learning-to-rank losses, may improve retrieval performance by better aligning model optimization with the retrieval task. Second, for codebook-based methods, the limited model-size scaling observed in our experiments suggests a need for larger training datasets and longer training schedules to fully realize their potential. Finally, although our inference scaling analysis focused on n-gram-based methods, codebook-based methods may also benefit from increased inference-time computation, such as using larger beam sizes—offering a valuable direction for extending the scope of inference scaling in generative retrieval.